\documentclass[fleqn,10pt]{wlscirep}
\title{Photon pair generation in hydrogenated amorphous silicon microring resonators}

\author[1,*]{Elizabeth Hemsley}
\author[1]{Damien Bonneau}
\author[2]{Jason Pelc}
\author[2]{Ray Beausoleil}
\author[1]{Jeremy L. O'Brien}
\author[1]{Mark G. Thompson}
\affil[1]{Quantum Engineering Technology Labs, H. H. Wills Physics Laboratory and Department of Electrical \& Electronic Engineering, University of Bristol, BS8 1FD, UK.
}
\affil[2]{ Hewlett-Packard Laboratories, 1501 Page Mill Rd., Palo Alto, CA 94304-1126 , USA}

\affil[*]{eh8423@bristol.ac.uk}

\usepackage{graphicx} 
\usepackage{subfig} 

\begin{abstract}
 We generate photon pairs in a-Si:H microrings using a CW pump, and find the Kerr coefficient of a-Si:H to be $3.73 \pm 0.25 \times 10^{-17}m^2/W$. By measuring the Q factor with coupled power we find that the loss in the a-Si:H micro-rings scales linearly with power, and therefore cannot originate from two photon absorption. Theoretically comparing a-Si:H and c-Si micro-ring pair sources, we show that the high Kerr coefficient of this sample of a-Si:H is best utilized for microrings with Q factors below $10^3$, but that for higher Q factor devices the photon pair rate is greatly suppressed due to the first order loss. 

\end{abstract}
\begin{document}

\flushbottom
\maketitle
\thispagestyle{empty}

\section*{Introduction}

Hydrogenated amorphous silicon (a-Si:H) has recently arisen as a photonics platform to contend with crystalline silicon (c-Si). a-Si:H can be deposited at low temperatures and so is easy to incorporate in the chip manufacturing processes, and retains many of the advantages of silicon based photonic structures - similar waveguide dimensions, CMOS compatibility and low loss guiding of telecom wavelengths. Additionally, the nonlinear properties of a-Si:H are known to be dependent on the manufacturing conditions, potentially giving a user freedom to design the material to his or her requirements.  a-Si:H has quasi-bandgap of $\sim1.7$eV, and experiences less two photon absorption (TPA) than c-Si, which has bandgap energy 1.1eV. Samples have been fabricated with Kerr coefficient several times larger than c-Si, overall giving a-Si:H a high nonlinear figure of merit. \cite{wang2012ultralow,lacava2013nonlinear,shoji2010ultrafast,grillet2012amorphous,gai2014negligible,kuyken2011nonlinear,matres2013high,narayanan2010broadband,pelc2014picosecond}. 

These properties make a-Si:H a potentially useful material for quantum optics applications, in particular photon pair generation via spontaneous four wave mixing (SFWM). In c-Si photon pairs can be generated by SFWM, but nonlinear losses such as TPA reduce the heralding efficiency of silicon sources for both microrings and waveguides. High heralding efficiency is a key requirement for a single photon source in linear optical quantum computing, and there are concerns that this inherent nonlinear loss in c-Si makes it unsuitable for large scale applications \cite{jennewein2011single,husko2013multi}.
 Conversely, a-Si:H waveguides have shown high photon pair generation efficiency without TPA \cite{clemmen2010generation,wang2014multichannel}, making a-Si:H attractive as a potential source platform. While waveguide sources are useful for many applications, microring based sources can generate spectrally pure single photons with small footprint and high efficiency \cite{clemmen2009continuous,harada2008generation, helt2010spontaneous}, and are important devices for quantum information processing. However, the field enhancement and long lifetime of light inside the microring means it is more susceptible to losses via free carrier absorption (FCA) than a waveguide, and so requires further investigation.

In this work we study the performance of a-Si:H microring resonators, in comparison to c-Si. We find that a-Si:H has a Kerr coefficient $\sim8$ times higher than c-Si, and that the main source of loss is linear with power, and so cannot originate from TPA. Instead, we model this loss by interband transitions facilitated by a mid-gap state, referred to here as sequential photon absorption (SPA) \cite{ito2014theoretical}, but also known as two stage absorption \cite{ikeda2007enhanced} and one photon absorption \cite{matsuda2009all}, among others. There is evidence that these mid-gap states exist in some a-Si:H samples, allowing non-instantaneous absorption of photons, but the origin of these states is not fully understood \cite{wathen2014non}. We assume that the main loss mechanism in the a-Si:H devices is free carrier absorption, analogous to c-Si, but where these free carriers are generated linearly with power. This type of loss was not observed in previous experiments with waveguide sources  \cite{clemmen2010generation,wang2014multichannel}, possibly because these experiments did not access the regime where mid-gap assisted FCA becomes significant in waveguide. Despite the high Kerr coefficient, the a-Si:H microrings studied here had an advantage over c-Si devices only for devices with low Q factor, due to the significant nonlinear loss at low power.

\subsection*{Devices and Theory}

\begin{figure}[ht]
	\centering
		\includegraphics[width=\textwidth]{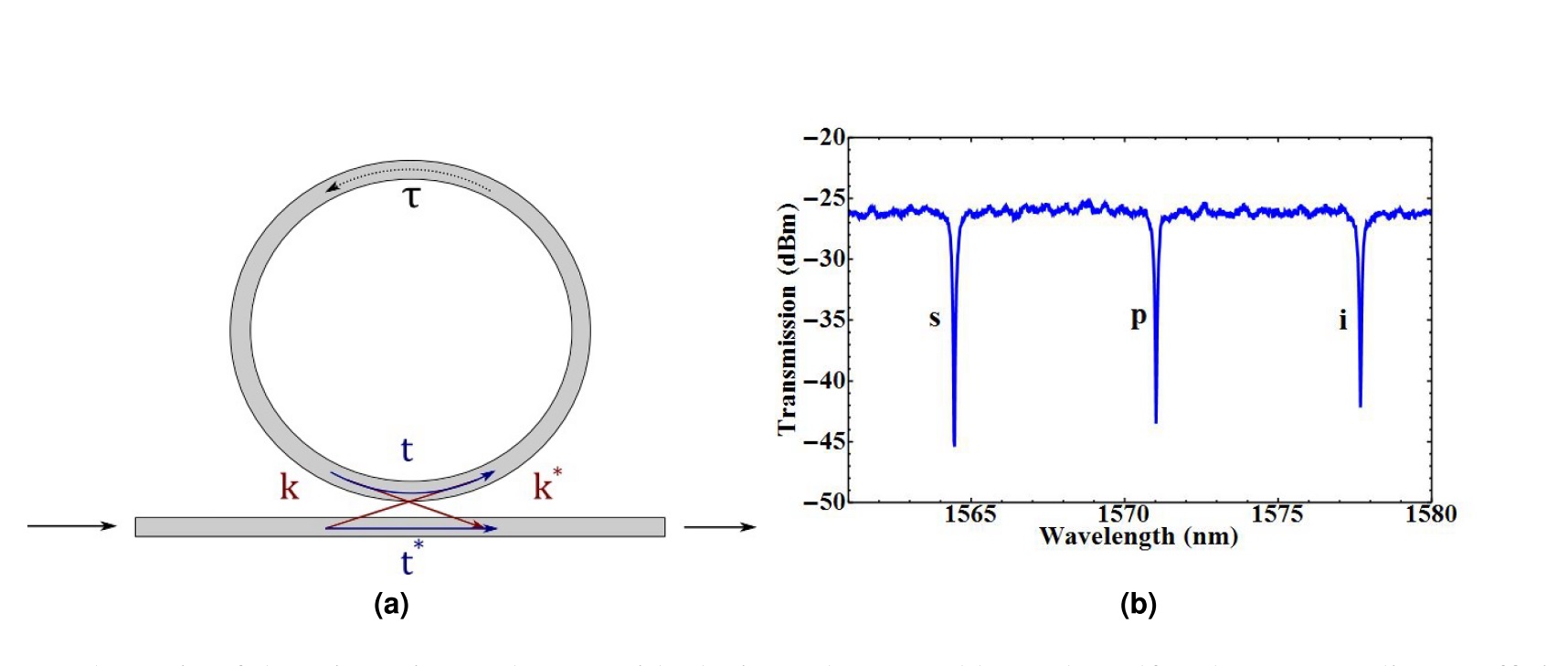}
		\caption{a) Schematic of the micro-ring and waveguide device, where t and k are the self and cross coupling coefficients, and $\tau$ the transmission per round trip. b) Example transmission scan through the device, with pump, signal and idler wavelengths indicated.}
	\label{fig:diags}
\end{figure}

The test structure was a a-Si:H micro-ring resonator coupled to a bus waveguide, illustrated in Figure \ref{fig:diags}a. We tested three such of these devices, with properties summarized in Table \ref{table:rings}. The chip was fabricated at HP Laboratories according to the recipe described in \cite{pelc2014picosecond}, and the rings were circular with diameter $27\mu m$, and waveguide effective area $0.3\mu m^2$. Light was coupled into the device using grating couplers with loss approximately 8dB each. 

 The ring resonator acts as a cavity, with a comb of resonances equally spaced in frequency. When light is is coupled into the ring and SFWM occurs, the signal and idler photons are generated at resonances either side of the pump wavelength, as shown Fig \ref{fig:diags}b. For a continuous wave (CW) pump, when the signal and idler photons are collected from a single pair of resonances the photon pair generation rate is given by \cite{helt2010spontaneous,azzini2012classical}

\begin{equation}
R=(P\gamma )^2 4 \frac{ Q^3 v^4_g}{\omega^3 L^2}(1\pm\sqrt{\Gamma})^3
\label{eq:Rlin}
\end{equation}

$P$ is the power in the bus waveguide, $L$ is the ring length, $Q$ is the Quality factor of the ring, and $v_g$ is the group velocity. The nonlinear coefficient $\gamma=\frac{n_2 \omega}{c A_eff}$, and $\Gamma$ is the normalised transmission through the waveguide when the light wavelength is at the resonant wavelength, known as the extinction ratio. When $\Gamma$=0 the ring is critically coupled and the loss per round trip $\tau$ is equal to the self coupling $t$. The sign of $\sqrt{\Gamma}$ is positive if the ring is over-coupled ($\tau < t$), and negative if the ring is under-coupled ($\tau > t$). SFWM is driven by the Kerr coefficient $n_2$, and while we assume device dimensions and properties are fixed, the Q factor of the ring is dependent on the loss in the ring \cite{Rabus2007}. If there is sufficient intensity of light in the ring, nonlinear losses increase the loss per round trip, subsequently decreasing the effective Q factor of the ring \cite{Guo2014Impact}. The extinction ratio is similarly influenced. These effects are commonly seen in silicon rings, as in drop in photon pair detection rate as coupled power increases \cite{engin2013photon,azzini2013stimulated}. \\

The power dependent Q factor is given by 

\begin{equation}
Q(P)=\frac{Q_0}{1+\frac{Q_o v_g}{\omega}\alpha_{NL}(P)}
\label{eq:Q(P)}
\end{equation}

where $\alpha_{NL}$ is the loss per metre in the ring due to nonlinear losses only. $Q_0$ is the Q factor of ring when the coupled power is zero. The extinction ratio $\Gamma$ also changes due to nonlinear losses, and can be described in terms of the Q factor \cite{Guo2014Impact}.

\begin{align}
	\frac{(1\pm\sqrt{\Gamma(P)})}{(1\pm\sqrt{\Gamma_0})}=\frac{Q}{Q_0}
	\label{eq:T(P)}
\end{align}

The extinction ratio $\Gamma$ also gives information about the coupling of the device. If a ring is over-coupled, $\Gamma$ will decrease with power, because the coupling condition becomes closer to critical coupling as the loss is increased. Conversely, if the ring is under coupled $\Gamma$ will increase with power. In this work, a-Si:H rings B and C were under-coupled, and a-Si:H ring A was over-coupled.

With a CW pump, the powers involved are generally too small for electronic state transitions via absorption of photons (such as TPA, 3PA etc) to significantly influence the loss in the ring. However, these processes generate free electrons in the conduction band which can absorb photons via intraband transitions, known as free carrier absorption (FCA). The number of photons absorbed in the creation of a free carrier is much lower than the number of photons absorbed by the free carriers themselves, because the free carrier can absorb several photons during its lifetime ($\sim 1ns$), and so we can neglect loss by TPA directly \cite{Yin2007}. 
Assuming all nonlinear loss in the ring is due to FCA only

\begin{align}
	\alpha_{NL}(P)=\sigma N
\end{align}

where N is the free carrier density in the ring, and $\sigma$ the FCA cross-section. When free carriers are generated via TPA, the free carrier population is governed by 

\begin{align}
	\frac{dN_{TPA}}{dt}=\frac{\beta_{TPA}}{2 h v_0 A_{eff}^2} (|F|^2 P)^2-\frac{N}{\tau_c}
\end{align}

where $|F|^2$ is the field enhancement in the ring, P is the power in the waveguide, and $\beta_{TPA}$ is the TPA coefficient. The bandgap of a-Si:H is typically larger than in c-Si, in part due to the presence of stable hydrogen bonds, which leads to reduced TPA coefficient in a-Si:H. However, SPA generates free carriers linearly with power, and in this case the free carrier density is given by

\begin{align}
	\frac{dN_{SPA}}{dt}=\frac{\beta_{SPA}}{h v_0 A_{eff}}|F|^2 P-\frac{N}{\tau_c}
	\end{align}

where $\beta_{SPA}$ is the SPA coefficient \cite{matsuda2009all}. Since we are operating in CW mode, the system is in a steady state where $\frac{dN}{dt}=0$. Depending on how the free carriers are generated, the Q factor is given by

\begin{align}
	Q_{TPA}(P)=\frac{Q_{0}}{1+\frac{Q_0 v_g}{\omega}\frac{\sigma\tau_c \beta_{TPA}  (|F|^2P)^2}{2hv_0 A_{eff}^2}}\\
	Q_{SPA}(P)=\frac{Q_{0}}{1+\frac{Q_0 v_g}{\omega}\frac{\sigma\tau_c \beta_{SPA} |F|^2P}{2hv_0 A_{eff}}}
	\label{eq:Qs}
\end{align}

where the field enhancement $|F|^2$ itself depends on the Q factor\cite{helt2010spontaneous}.

\begin{align}
	|F|^2=\frac{2vQ}{L \omega} (1\pm\sqrt{\Gamma})
	\label{eq:F}
	\end{align}

Using Equations \ref{eq:T(P)}, \ref{eq:Qs} and \ref{eq:F}, the power-dependent Q factors for materials with TPA and SPA generated free carriers respectively are given by

\begin{align}
Q_{TPA}(P)=\frac{2Q_{0}}{1+\sqrt{1+\frac{8 (1\pm\sqrt{\Gamma_0})^2 v_g^3 \sigma\tau_c \beta_{TPA}}{L^2hv_0 A_{eff}^2\omega^3}Q_0^3 P^2}}
\label{eq:Q_{TPA}}\\
Q_{SPA}(P)=\frac{2Q_{0}}{1+\sqrt{1+\frac{4 (1\pm\sqrt{\Gamma_0}) v_g^2 \sigma\tau_c \beta_{SPA}}{L hv_0 A_{eff}\omega^2}Q_0^2 P}}
\label{eq:Q_{SPA}}
\end{align}

The scaling of the Q factor with power can be used to determine which nonlinear loss mechanism is present in the material.
Using Equation \ref{eq:Rlin} with \ref{eq:Q_{SPA}} or  \ref{eq:Q_{TPA}}, the photon pair generation rate is then given by,

\begin{align}
	R=\frac{4 (1\pm\sqrt{\Gamma_0})^3 (\gamma P )^2 v^4 Q(P)^6}{L^2 Q_0^3 \omega^3}
	\label{eq:Rate}
\end{align}

Where $Q(P)$ can be given by $Q_{SPA}(P)$ or $Q_{TPA}(P)$.

\section*{Power-Dependent Q Factor}

\begin{figure}[ht]
	\centering
		\includegraphics[width=\textwidth,]{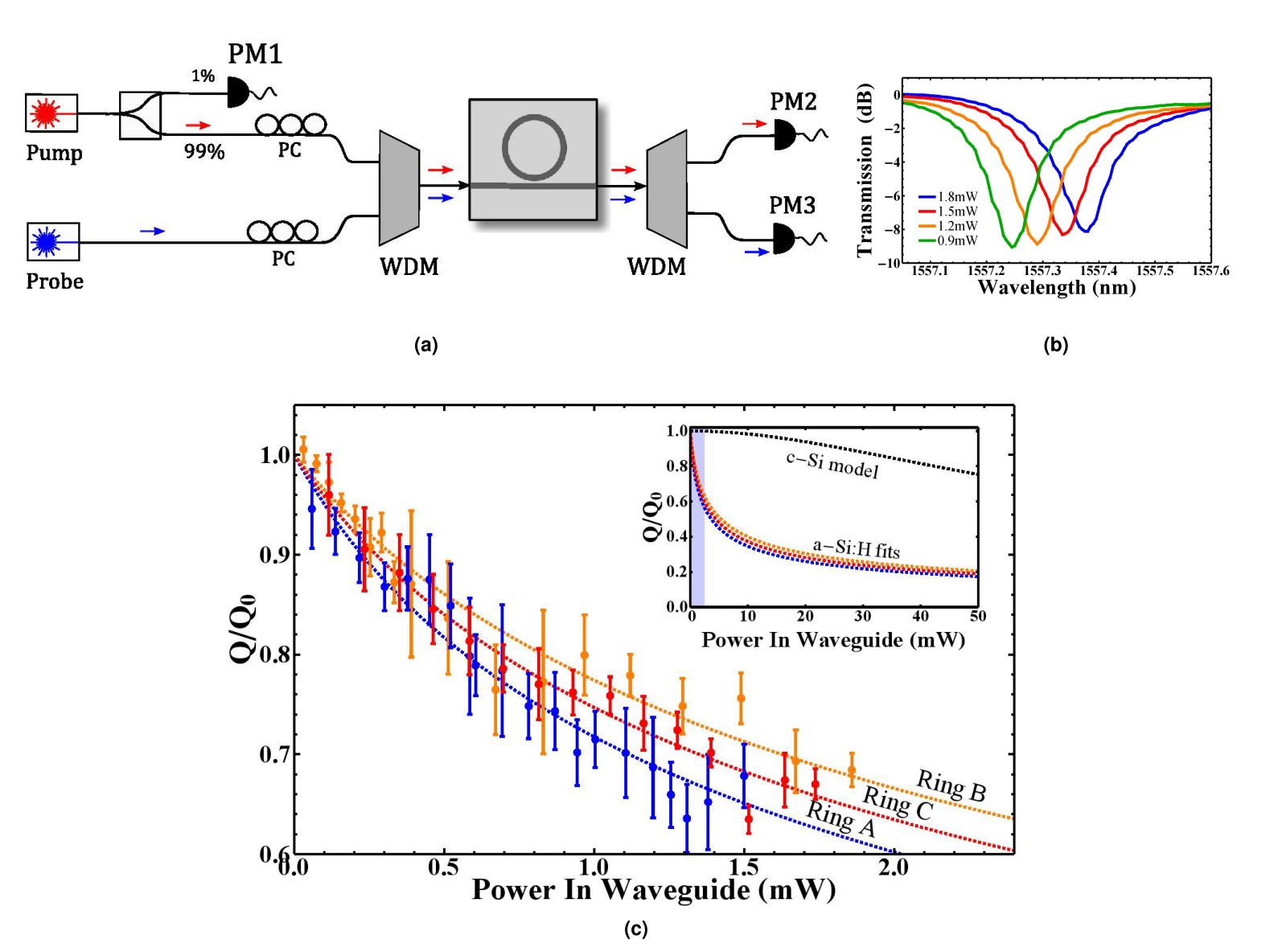}
		\caption{a) Experimental setup to measure the power scaling of the Q factor, $Q/Q_0$. b) Example probe transmission scans for a-Si:H ring B, for various coupled pump powers.  c) $Q/Q_0$ for the a-Si:H devices, fit to equation \ref{eq:Q_{SPA}}, error bars are determined by averaging over several data sets. The subplot shows the extrapolated fits for $Q/Q_0$ compared to a model for an equivalent c-Si micro-ring, based on equation \ref{eq:Q_{TPA}}. The region of the main plot is indicated by the shaded area in the sub-plot.}
	\label{fig:Q}
\end{figure}

The experimental set-up used to measure the Q factor with coupled pump power is shown in Fig. \ref{fig:Q}a. A bright CW pump  was coupled into the chip, and the CW wavelength was tuned until it was at one of the ring resonance wavelengths. A weak CW probe (\textless 0.1mW in the waveguide) then performed a wavelength scan in the region of an adjacent resonance. A 200GHz wavelength division multiplexer (WDM) from Opneti is used to combine the pump and probe beam before the chip, with an identical device after the chip to separate the pump and probe wavelengths. Monitoring the out-coupled probe light at PM3 gives the device transmission spectrum, which we use to extract the Q factor for a given pump power. Power meters PM1 and PM2 are used to monitor the coupling into the chip. The pump CW light was generated by a tunable fibre coupled laser, amplified by an erbium-doped fibre amplifier, and propagated through a polarization controller and variable attenuator, using a 99:1 in-fibre beam-splitter to monitor input power. The probe light was generated in a similar CW tunable laser, and also passed through a polarization controller, so that both beams were maximally coupled to the waveguide polarization mode. A peltier was used to control the temperature of the chip, which was adjusted to optimize the ring resonance wavelength with respect to the DWDM channel wavelengths.

The Q factor for the a-Si:H rings are shown in Figure \ref{fig:Q}c. The Q factor decreases even at low powers with a linear power dependence, which we attribute to SPA based loss. The results were fit using equation \ref{eq:Q_{SPA}}, and we extract the parameter $\sigma\tau\beta_{SPA}$, giving $2.44 \pm 0.223$, $2.04 \pm 0.200$ and $2.13 \pm 0.176 \times 10^{-27} ms$ for rings A, B and C respectively. If there is any loss due to TPA generated free carriers present in the a-Si:H devices, it is negligible compared to the loss due to SPA generated free carriers, for the powers measured here. Since TPA is quadratic with power there exists a power threshold where TPA becomes dominant over SPA, so we cannot assume TPA is not present in this sample of a-Si:H, merely that we have not yet reached the power levels where it becomes important.

To compare the behaviour of these rings to that expected for crystalline silicon, we modelled the power dependent Q factor for a c-Si device with $Q_0=11070$, (the average value of the three a-Si:H rings) and identical length to the a-Si:H devices, but where the main loss mechanism was due to TPA driven FCA, with TPA coefficient $8.4\times 10^-12 W/m^2$ \cite{dinu2003third}. The result, displayed in the subplot in figure \ref{fig:Q}c, show that the Q factor for the c-Si micro-ring is much more stable than a-Si:H, and does not experience any significant decrease until over 40mW of coupled power.

\section*{Photon Pair Generation}

To measure the rate of photon pairs generated via SFWM in the ring a CW laser was amplified and filtered to remove background noise from the amplifier and the wavelength tuned to the ring resonance wavelength. The signal and idler photons generated on resonance were separated from the residual pump light using a WDM after the chip and detected using super-conducting nanowire single photon detectors (SNSPDs). Coincidence counting hardware was used to obtain the rate of coincidence detections between the two SNSPDs, and accidental coincidence counts were removed. The rate of coincidences originating from photons generated in the waveguide was negligible in the devices.

\begin{figure}[ht]
	\centering
		\includegraphics[width=0.70\textwidth,trim = {0.5cm 0.5cm 0cm 0.5cm},clip]{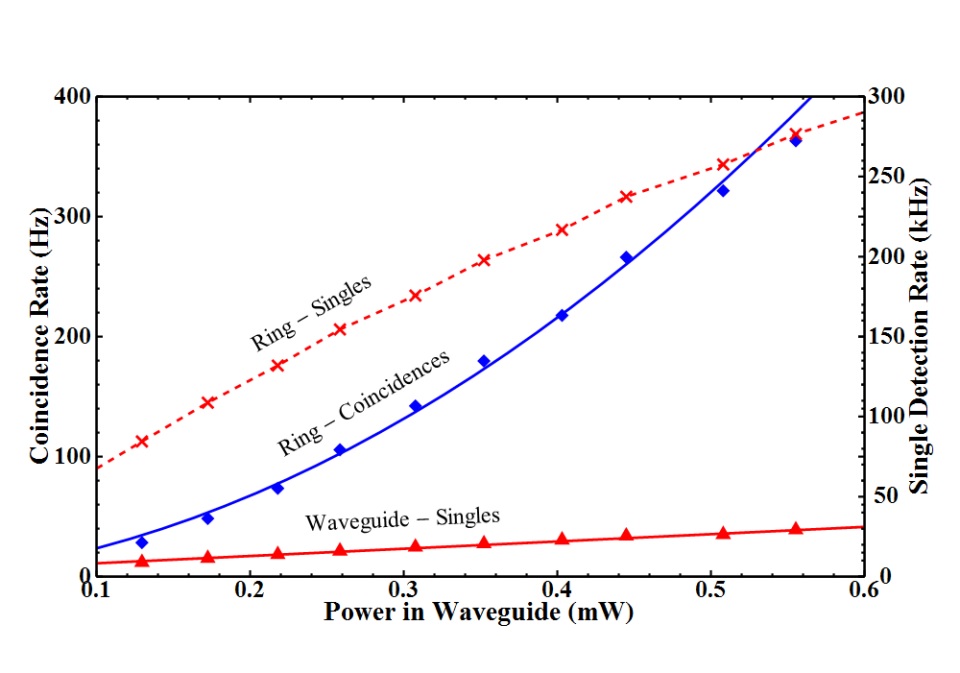}
		\caption{Coincidence and singles rate for a-Si device A. Red crosses represent the detection rate of a single detector when the CW	laser is tuned to the ring resonance wavelength, red triangles the detection rate when the CW laser is not tuned into the ring resonance and propagates through the waveguide only, and blue diamonds give the rate of coincidence counts between the two detectors when the CW is tuned into the ring resonance. The ring singles are joined with the red dashed line to help guide the eye, the red solid line is a linear fit to the waveguide singles, and the blue solid line is a quadratic fit to the coincidence count data. Error bars were smaller than the symbols. }
	\label{fig:raman}
\end{figure}

The Raman spectrum of a-Si:H has a peak centred around 480 $cm^{-1}$, corresponding to a wavelength shift of $\approx$ 115nm for 1550nm photons, but the width of the peak is typically broader than in c-Si and tails off into lower wavenumbers due to the presence of hydrogen bonds \cite{iqbal1982raman,beeman1985structural,brodsky1977infrared}. This broad peak allows pump photons to be scattered to energies corresponding to the signal and idler wavelength. The single detection rates measured were dominated by these spontaneous Raman scattered photons, illustrated by Figure \ref{fig:raman}. The rate of single detections is much higher when the pump wavelength is tuned to the ring resonance wavelength, but does not have a quadratic relationship with power, indicating that these counts are due to photons generated in the ring, but not by SFWM. Due to the large contribution of accidental counts due to these Raman scattered photons, the CAR for devices was low, reaching a maximum of 10.6, 12.3 and 9.3, at powers 1.1, 1.7 and 3.5mW for rings A, B and C respectively. 

Using the rate of the single counts at each detector, and the rate of coincidences, it is possible to remove the loss of each channel and obtain the absolute rate of photons generated in the ring, following the method outlined in \cite{engin2013photon}. However this method requires that the single detection rates have a quadratic component to their relationship with power, allowing detection events originating from pump leakage or Raman scattered photons to be removed, and so could not be used directly to calculate the absolute rate for the a-Si:H devices. Instead, we used a c-Si micro-ring with high CAR to find the detector efficiencies, and then combined with measured channel losses to determine the absolute photon pair generation rate for the a-Si:H devices.

\begin{figure}[ht]
	\centering
		\includegraphics[width=\textwidth]{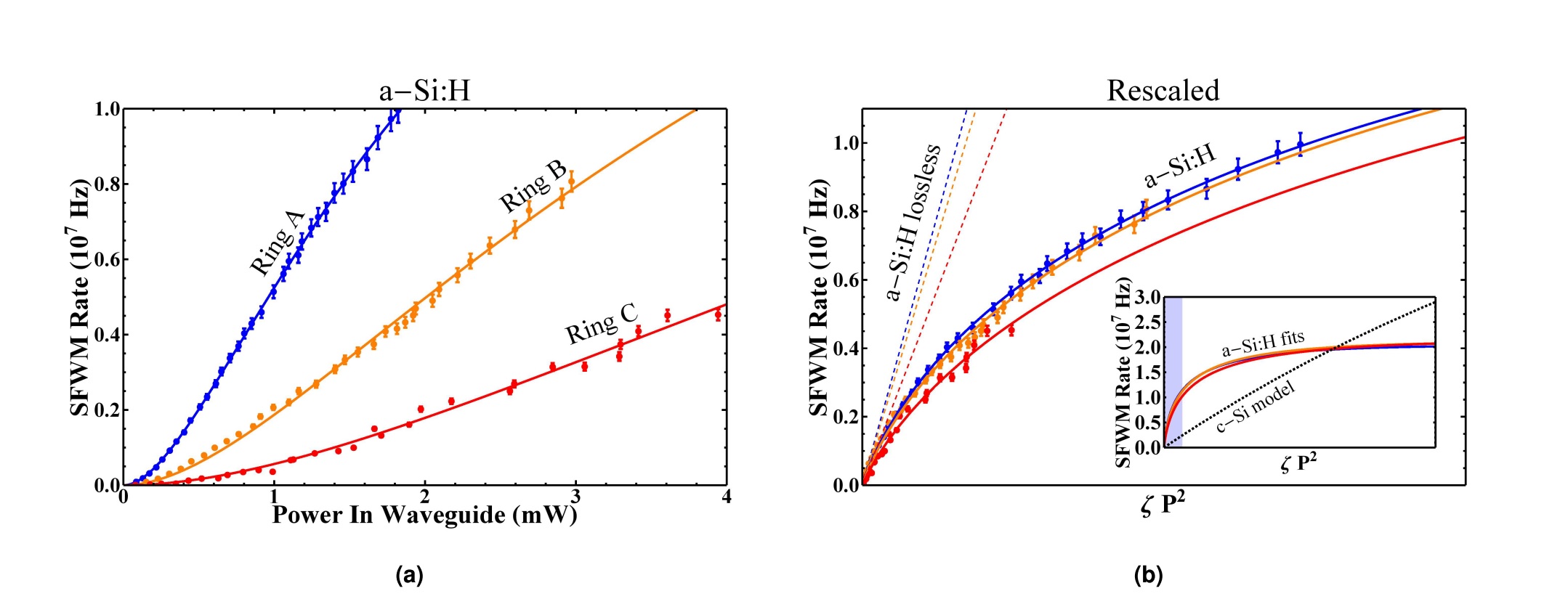}
		\caption{a) Calculated SFWM photon pair generation rate for a-Si:H devices. Error bars due to Poissonian error in measured count rate. b)SFWM photon pair generation rate with x-axis rescaled. Solid lines are fit to the data using equation \ref{eq:Rate}, and dashed lines indicate the case without nonlinear loss, with slope proportional to $n_2^2$. The subplot compares the extrapolated fits from the a-Si:H data to a model for a c-Si micro-ring. The region of the main plot is indicated by the shaded area in the sub plot.}
	\label{fig:counts}
\end{figure}

The generation rates for the rings are shown in Figure \ref{fig:counts}a, fit to equation \ref{eq:Rate} using the equation for SPA driven loss given by \ref{eq:Q_{SPA}}.  For the a-Si:H devices we measured the Kerr coefficient to be $n_2$ to be $3.98 \pm 0.03$, $3.82\pm 0.07$ and $3.39\pm 0.1 \times 10^{-18} m^2/W$ for rings A, B and C respectively, approximately 8 times larger than the literature value for c-Si. We measured the SPA parameter $\sigma\tau\beta_{SPA}$ to be $0.87\pm 0.01$, $0.55 \pm 0.05$ and $0.53 \pm 0.06 \times 10^{-27} ms$ for rings A, B and C. The discrepancy between these values and the previous section likely arise from the increased stability during the measurement (see Supplementary Information). 

In in Figure \ref{fig:counts}b we rescale the data using $\zeta=4c^2Q^3(1\pm \sqrt{\Gamma_0})^3(\omega L^2n^4A_{eff}^2)^{-1}$ so that the loss-less SFWM rate $R=n_2^2 \zeta P^2$. This compensates for the differences in Q factors and extinction ratios between the rings, and gives a good overlap for the data collected from each ring. Again we include a model for a c-Si ring with Q factor 11070, which demonstrates that while initially the a-Si:H rings are more efficient at photon pair generation, at higher powers the count rate is significantly suppressed compared to c-Si. 

\section*{Theoretical Comparison}

\begin{figure}[ht]
\centering
	\includegraphics[width=1\textwidth,trim = {1cm 1cm 3cm 1cm},clip]{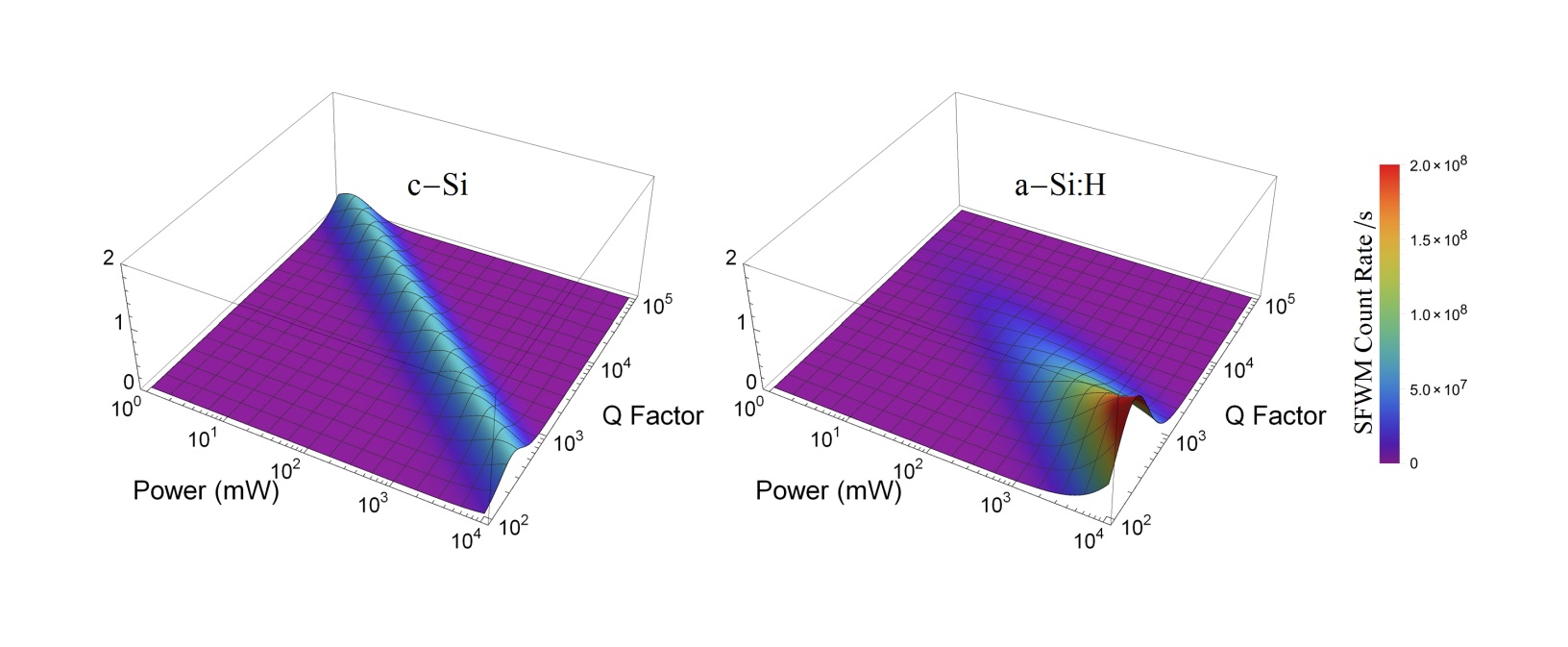}
			\caption{Comparison of photon pair generation rate from c-Si and a-Si:H microrings with Q factor and input power.}
	\label{fig:com}
\end{figure}

We theoretically compare the performance of identical c-Si and a-Si:H microrings using the parameters and loss mechanisms determined in these experiments, for devices with a range of Q factors, shown in Figure \ref{fig:com}. The SFWM rate for c-Si ring is modelled using $n_2=0.45\times10^{-17}m^2/W$, with the main loss mechanism due to absorption via TPA generated free carriers, with TPA coefficient $\sigma\tau\beta_{TPA}=8.4 \times 10^{-40}m^3s/W$. The SFWM rate for the a-Si:H devices is modelled using the average value for the SPM coefficient determined experimentally in the previous sections, $\sigma\tau\beta_{SPA}=1.1 \times 10^{-27} ms$, with Kerr coefficient $n_2=3.73\times 10^{-17} m^2/W$. Since in this model we examine higher powers than those used experimentally, we also include loss via TPA, with $\sigma\tau\beta_{TPA}=1.4 \times 10^{-40}m^3s/W$ \cite{pelc2014picosecond}, although this loss is not significant when compared to SPA based losses. In other respects, the two devices are identical in terms of extinction ratio and length, although c-Si is modelled with $A_{eff}=0.2\times 10^{-12}$. Of course, while the ring properties, $Q, L, A_{eff}$ and $\Gamma$ are not entirely independent, and do additionally depend on the material properties, this model gives an indication of how c-Si performs in general compared to a-Si:H.

For c-Si, as both loss and photon pair generation rate scale quadratically with input power, the maximum possible SFWM rate from the device is constant, with the device becoming more efficient for high Q factor devices. For our sample of a-Si:H, it is clear that the SFWM rate for high Q factor devices is greatly suppressed by nonlinear losses. a-Si:H can generate higher photon pair rates that c-Si, but only for devices with Q factor below $\sim 10^3$, at the expense of injecting much higher powers.

\section*{Conclusion}

We measured the power dependent Q factor for a number of a-Si:H micro-ring resonators, and find that the loss in the ring increases linearly with coupled power, compared to quadratically scaling TPA based loss typically seen in c-Si. We suggest that the loss in a-Si:H is due to the presence of mid-gap states, which allow free carrier generation by bridging the electronic bandgap and allowing a stepping stone to the conduction band. We then generated photon pairs in the devices via SFWM with a CW pump, and confirm that a-Si:H has a higher Kerr coefficient than c-Si, but show that the photon pair generation rates are limited due to this loss. We also find there is a large contribution to the count rate from Raman scattered photons, which decrease the signal to noise ratio of the a-Si:H source. The number of Raman scattered photons may be minimized by operating the device at cryogenic temperatures, or choosing manufacturing conditions to reduce the number of H bonds which broaden the Raman peak. 

By theoretically comparing c-Si and a-Si:H microrings, we show low Q factor a-Si:H devices have the potential to generate much higher rates of photon pairs than c-Si, at the expense injecting much more pump power. However, c-Si microrings generate more photon pairs per second than equivalent a-Si:H microrings for $ Q > 10^3$ devices, because of the significant first order nonlinear loss observed in a-Si:H. We suggest that this loss arises from the presence of mid-gap states, in a process known as sequential photon absorption. Since the properties of a-Si:H are influenced by the manufacturing conditions, optimising the fabrication method to minimize the number of mid-gap states would be greatly beneficial, and further research into the origin of these states is needed. While this type of loss seems to be problematic for a single photon sources, the low power generation of free carriers in a-Si:H may be of interest to the classical photonics community, by allowing ultra-low power optical modulation \cite{matsuda2009all}.

With a pulsed pump, we suggest that under some circumstances SPA may be less detrimental than TPA to the source heralding efficiency. SPA based loss only depends on the average power of the pump, while with loss originating with TPA the peak power is the key factor. Additionally, where SPA is dominant, loss due to cross-two photon absorption (simultaneous absorption of a pump photon with a signal or idler photon) is not an issue. Another feature of SPA is that the process is non-instantaneous, since the mid-gap states have a non-negligible lifetime. In the CW regime this has no effect, but using a short pulsed pump could further reduce loss from SPA generated free carriers. It is advantageous to use a-Si:H for waveguide based sources, and low Q factor a-Si:H microrings could be useful for applications where short pulses with high peak power are required, although further study is required to determine whether operating in this regime we could achieve better heralding efficiencies than c-Si in general.

\bibliography{literature}

\section*{Acknowledgements}

This work was supported by the Engineering and Physical Science Research Council (EPSRC, UK), the European Research Council, the Bristol Centre for Nanoscience and Quantum Information, the European FP7 project BBOI. J.L.OB. acknowledges a Royal Society Wolfson Merit Award and a Royal Academy of Engineering Chair in Emerging Technologies. M.G.T. acknowledges support from an Engineering and Physical Sciences Research Council (EPSRC) Early Career Fellowship. We also thank Gary Sinclair for many useful discussions.

\section*{Author contributions statement}
This paper was written by E.H., D.B. and M.G.T. The experiment was conceived by E.H., D.B., M.G.T. and J.P. The experiments and data analysis were conducted by E.H., and J.P. designed and fabricated the device. The experiment was motivated by R.B. and J. L. O'B.  All authors reviewed the manuscript. 

\section*{Competing financial interests}
The author(s) declare no competing financial interests.

	\begin{table}[ht]
\centering
\begin{tabular}{l||c|c|c||}
&\multicolumn{3}{c||}{a-Si:H}\\
	\hline\
	&A&B&C\\
	\hline
	Quality Factor, Q&10800&12700&9700\\
	Extinction Ratio, $\Gamma$ (dB)&13&11&9\\
	\hline
\end{tabular}
	\caption{Device Parameters}
	\label{table:rings}
	\end{table}

\end{document}